\documentclass[twocolumn,amsmath,superscriptaddress,amssymb,nopacs,nofootinbib,prb]{revtex4-2}

\usepackage[usenames,dvips]{color}
\usepackage{graphicx}
\usepackage{dcolumn}
\usepackage{bm}
\usepackage{mathtools}
\usepackage{epstopdf}
\usepackage{esint}
\usepackage{xcolor}
\usepackage{dsfont}
\usepackage{nicefrac}

\usepackage[colorlinks=true,citecolor=magenta,linkcolor=magenta, urlcolor=magenta]{hyperref}

\newcommand{\bea}{\begin{eqnarray}}
\newcommand{\eea}{\end{eqnarray}}
\newcommand{\bt}{\textbf}

\newcommand{\phd}{\phantom{\dag}}

\newcommand{\noi}{\noindent}
\newcommand{\no}{\nonumber}

\begin{document}
\def\v#1{{\bf #1}}

\title{Nonreciprocal Equilibrium Josephson Effect of Arbitrary Periodicity\\  from Poor Man's Majorana Zero Modes}

\author{Panagiotis Kotetes}
\email{kotetes@baqis.ac.cn}
\affiliation{Beijing Academy of Quantum Information Sciences, Beijing 100193, China}
\affiliation{CAS Key Laboratory of Theoretical Physics, Institute of Theoretical Physics, Chinese Academy of Sciences, Beijing 100190, China}

\author{Merc\`e Roig}
\affiliation{Niels Bohr Institute, University of Copenhagen, DK-2100 Copenhagen, Denmark}

\author{Brian M. Andersen}
\affiliation{Niels Bohr Institute, University of Copenhagen, DK-2100 Copenhagen, Denmark}

\vskip 1cm

\begin{abstract}
We show that the Josephson diode effect becomes possible when two coupled antiferromagnetic dimers of point-like magnetic adatoms are deposited on top of a Rashba superconductor. The degree of nonreciprocity is substantial when the arising Yu-Shiba-Rusinov (YSR) bound states approach zero energy. In this limit, these states behave as weakly coupled poor man's Majorana (PMM) excitations. This PMM regime is accompanied by highly dispersive and phase-bias asymmetric Andreev bound state dispersions. In turn, these result in a nonreciprocal Josephson current, whose diode efficiency can be controlled by varying the geometric details of the adatom's spatial configuration. In addition, thanks to spin-triplet pairing terms mixing different YSR states, the Josephson current can possess any periodicity in equilibrium, including $4\pi$. Our work opens the door to observing and harnessing Majorana be\-ha\-vior in currently experimentally accessible topologically trivial systems.
\end{abstract}

\maketitle

\textit{\bt{Introduction} -} Recent experiments~\cite{Ando2020,Lyu2021,Diez2023,Yun2023,Lin2022,Baumgartner2022,Strambini2022,Wu2022,Bauriedl2022,Hou2023,Pal2022,Turini2022,Trahms,Du2023,Gupta2023,Gutfreund2023,Narita2023,Margineda2023} have unveiled a new promising functionality of Rashba spin-orbit interaction (SOI). This is, to me\-dia\-te the superconducting diode effect~\cite{Daido2022prl,Daido2022prb,Ilic2022,Yuan2022,He2022,Zhang2022,Scammell2022,Souto,Legg2022,Davydova2022,Tanaka2022,BoLu2023,Karabassov2022,Steiner2023,Kochan2023,Costa2023,MeyerDiode,LevchenkoDiode,LeggMZM,QFSun,Cayao2023,Liu2023,RoigDiode,SoutoDanon,Debnath,Fracassi,BanerjeeDiode,Chakraborty}, which ma\-ni\-fests itself in a polarity-dependent critical supercurrent~\cite{Rikken,Wakatsuki,Tokura2018,Nadeem2023}. For the diode effect to take place, time-reversal symmetry (TRS) is also required to be violated. In the early diode experiments TRS was broken by means of a spatially-uniform in-plane magnetic field $\bm{B}_{\rm in}=(B_x,B_y)$~\cite{Ando2020,Baumgartner2022}. Since then, a plethora of alternative routes to rea\-li\-ze superconduc\-ting diodes have been proposed. Amongst these, one possi\-bi\-li\-ty is to couple the superconductor (SC) to an out-of-plane spatially-nonuniform magnetization $M_z(x,y)$~\cite{RoigDiode}. As first brought forward in Ref.~\onlinecite{KotetesAnatomy}, the presence of the Rashba SOI renders the magnetization gradients $(\partial_xM_z,\partial_yM_z)$ symmetry-equivalent to $(B_x,B_y)$, thus, also enabling nonreciprocal Josephson transport. Notably, in contrast to the uniform magnetic field, tailoring the spatial profile of the magnetization allows us to control the current distribution in space and provides more knobs for optimizing the diode efficiency. Our previous work discussed two different scena\-rios, i.e., a uniform gradient and a ferromagnetic domain wall~\cite{RoigDiode}. In both cases, the diode effect exclusively stemmed from bulk electrons, and followed from the analysis of Ref.~\onlinecite{KotetesAnatomy}.

\begin{figure}[t!]
\begin{center}
\includegraphics[width=1\columnwidth]{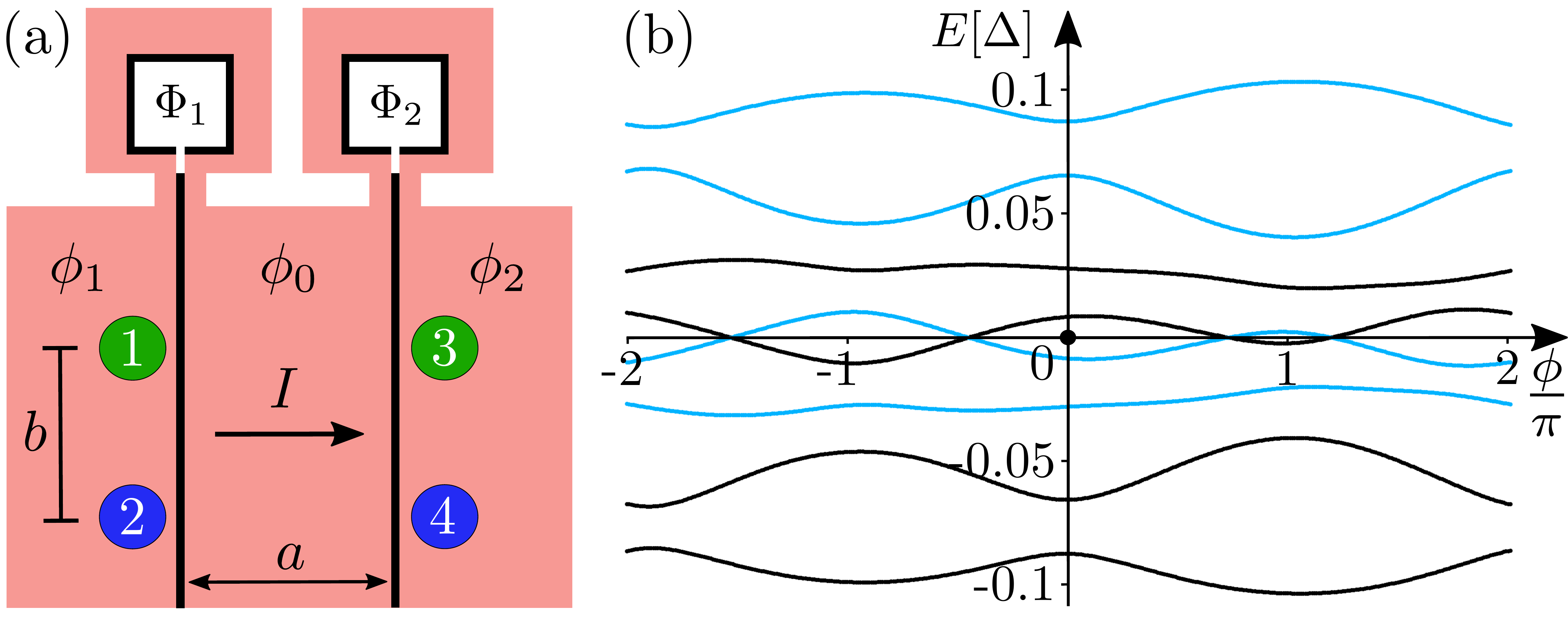}
\end{center}
\caption{(a) Sketch of the proposed two-AFM-dimer setup. Each AFM dimer is deposited on top of a Rashba SC (left and right). These SCs are kept at a phase difference $\phi=\phi_1-\phi_2$, with $\phi_1=\phi_0-\Phi_1$ and $\phi_2=\phi_0+\Phi_2$, where $\Phi_{1,2}$ denote flux biases expressed in units of phase. Here, $\phi_0$ is the phase of the middle Rashba SC. The spin moments of the ``1,\,3" (``2,\,4") adatoms are pointing out of (in to) the page. (b) Typical YSR dispersions obtained for (a) when $(\phi_1+\phi_2)/2=\phi_0$. A  $4\pi$-periodic Josephson diode effect appears even in equilibrium.}
\label{fig:Figure1}
\end{figure}

Here, we explore the superconducting diode effect which arises from magnetization gra\-dients that are instead generated from point-like adatoms embedded in a two-dimensional Rashba SC whose phase field possesses a spatially varying profile. In Fig.~\ref{fig:Figure1}(a) we depict the mi\-ni\-mal setup which relies on coupled pairs of antiferromagnetic (AFM) adatoms and can lead to a superconducting diode effect. This consists of two coupled AFM adatom dimers, with their magnetic moments poin\-ting out of the plane. Throughout, we focus on the nonreciprocal Josephson current which solely results from the Yu-Shiba-Rusinov (YSR) states induced by the adatoms~\cite{YSRS}.

Our analysis reveals that the diode effect becomes substantial when the energy of the YSR states is near zero. In fact, the combination of spin-singlet pairing and Rashba SOI ensures that the zero YSR states behave si\-mi\-lar to Majorana zero modes (MZMs), albeit that they have a fine-tuned rather than a to\-po\-lo\-gi\-cal origin~\cite{Leijnse,Tsintzis2022,Tsintzis2024,SoutoAguado}. Nonetheless, we show that these so-called poor man's Majoranas (PMMs) endow the Josephson current with characteristics that are typical to MZMs~\cite{TanakaPWave,Sengupta,KitaevUnpaired,FuKaneJ,Tanaka2009,Jiang2011,San-Jose2012,KotetesTI,Pekker,CarloJ,Sticlet2013,Cayao,Houzet,Levchenko,KMC,MKC,Sakurai,Wang2022,Javed2023}. Spe\-ci\-fi\-cal\-ly, we find that the current $I(\phi)$ can be both nonreciprocal and $4\pi$-periodic in terms of the phase bias $\phi=\phi_1-\phi_2$, with $\phi_{1,2}$ the phases felt by the two dimers. Notably, the $4\pi$-periodicity arises even in equilibrium, due to the phase $\phi_0$ of the middle SC in Fig.~\ref{fig:Figure1}(a)~\cite{Jiang2011,KotetesTI,Houzet,Levchenko,KMC,MKC,Sakurai}. In the remainder, we explore the Josephson current upon varying the intra-dimer distance and the exchange coupling strength of electrons and magnetic moments.

\textit{\bt{Modeling the system} -} The Rashba SC is described using the continuum Bo\-go\-liu\-bov - de Gennes (BdG) Hamiltonian:
\begin{align}
\hat{H}_0(\bm{p})=\tau_z\big[\varepsilon(p)+\upsilon\big(p_x\sigma_y-p_y\sigma_x\big)\big]+\Delta\tau_x\,,\label{eq:BulkHam}
\end{align}

\noi where $\bm{p}$ is the momentum, $\sigma_{x,y,z}$ denote the spin Pauli matrices, $\upsilon\geq0$ is the Rashba SOI strength, and $\Delta\geq0$ defines the spin-singlet pairing gap. In addition, we employ the energy dispersion $\varepsilon(p)=\big(p^2-p_F^2\big)/2m_*$, with $m_*$ the effective mass, $p_F$ the Fermi momentum of the system for $\upsilon=\Delta=0$, and $p=|\bm{p}|$. The above ``single"-particle Hamiltonian is expressed in the Nambu space spanned by electrons with spin up and down, along with their hole partners after acting on them with the time-reversal operator $\hat{\cal T}=i\sigma_y\hat{\cal K}$, where $\hat{\cal K}$ effects complex conjugation. Operators in Nambu space are represented with the help of the Pauli matrices $\tau_{x,y,z}$. For convenience of notation, the respective unit matrices in spin, Nambu, and other spaces are omitted throughout, and we set $\hbar=1$ for the reduced Planck constant.

We now proceed with the adatoms. These are assumed to have a spatial extent which is smaller than the Fermi wavelength $\lambda_F=2\pi/p_F$. For this reason, we model them as point-like objects. Moreover, their magnetic moments are considered classical and orien\-ted out of the system's plane. In their presence, the coordinate space BdG Hamiltonian takes the additional contribution $\hat{V}(\bm{r})=\sum_n\hat{V}_n\delta(\bm{r}-\bm{R}_n)$, where $\bm{R}_n$ are the positions of the adatoms in coordinate space. The terms $\hat{V}_n=JS_n\sigma_z$ denote the exchange interactions of electrons with the localized spins $S_n=\pm1$, where $J$ is the coupling strength.

\textit{\bt{Effective YSR Hamiltonian} -} The addition of the classical magnetic adatoms induces YSR bound states in the underlying Rashba SC~\cite{YSRS}. The YSR energy spectrum can be obtained from the poles of the so-called T-matrix~\cite{BalatskyRMP}. After the results of prior stu\-dies~\cite{Nakosai,Pientka,Heimes,Brydon,Heimes2}, pro\-jec\-ting the T-matrix onto two adatom positions $\bm{R}_{n,m}$ and restricting to energies which are much smaller than $\Delta$, leads to the YSR coupling Hamiltonian elements:
\begin{align}
\hat{H}_{nm}=\left[\tau_x+\big(\pi\nu_F\hat{V}_n\big)^{-1}\right]\delta_{nm}+\big(1-\delta_{nm}\big)\hat{g}_{nm},
\label{eq:YSRcouplings}
\end{align}

\noi with the energy expressed in units of $\Delta$. In the above, $\delta_{nm}$ is the Kronecker delta and $\nu_F=m_*/2\pi$ is the density of states per spin at the Fermi level when $\upsilon=\Delta=0$. We also defined the off-diagonal Hamiltonian elements $\hat{g}_{nm}\equiv\hat{g}(\bm{R}_{nm})$, which are functions of the relative position vector $\bm{R}_{nm}=\bm{R}_n-\bm{R}_m\equiv R_{nm}\big(\cos\theta_{nm},\sin\theta_{nm}\big)$ with $R_{nm}=|\bm{R}_{nm}|$. In the remainder, we assume that $p_FR_{nm}>\pi$. In this regime, previous results~\cite{Heimes2} yield:
\begin{align}
\hat{g}(\bm{R})=f(R)e^{-i\zeta(R)\tau_y}\tau_x\Big\{\cos[\eta(R)]-i\sin[\eta(R)]\sigma_ye^{i\theta\sigma_z}\Big\}
\end{align}

\noi where we momentarily suppressed the indices $n$ and $m$.

We observe the presence of three new parameters. The first is $f(R)={\rm Exp}(-R\Delta/\upsilon_F)\sqrt{2/(\pi p_FR)}$, with the Fermi velocity $\upsilon_F=p_F/m_*$, and controls the overall strength of the YSR coupling. Next, is the variable $\eta(R)=m_*\upsilon R$ which results from the Rashba SOI and controls the ratio between spin-singlet and spin-triplet hopping terms. Lastly, the parameter $\zeta(R)=p_FR-\pi/4$ tunes the ratio between electron and pair hoppings, which enter in the Hamiltonian with terms $\propto\tau_z$ and $\propto\tau_x$, respectively. From now on, we choose to express all lengthscales in units of $\lambda_F$. This allows us to write the above coefficients solely in terms of the three dimensionless quantities: $R/\lambda_F>1/2$, $\Delta/E_F\ll1$ and $2m_*\upsilon/p_F\ll1$. In the remainder, we consider $\Delta/E_F=2m_*\upsilon/p_F=0.1$ and $a/\lambda_F=2.8$ for all the nu\-me\-ri\-cal calculations, while results for other parameter values are provided in Ref.~\onlinecite{Sup}.

We now proceed by infer\-ring the Hamiltonian which describes the coupled YSR states obtained in the setup of Fig.~\ref{fig:Figure1}(a). To express the Hamiltonian in the two-state space of adatoms which comprise a single dimer, we employ the Pauli matrices $\rho_{x,y,z}$, with $\rho_z=\pm1$ labelling adatoms ``1,3" and ``2,4" in Fig.~\ref{fig:Figure1}(a). When the two identical AFM dimers (``1-2" and ``3-4") are decoupled and their constituent adatoms are separated by a distance $b$, they are both dictated by the intra-dimer Hamiltonian:
\bea
\hat{H}_{\rm intra}&=&\frac{1+\rho_z}{2}\big(\tau_x+ SM\sigma_z\big)+\frac{1-\rho_z}{2}\big(\tau_x- SM\sigma_z\big)\no\\
&&+\Big[\rho_+f_be^{-i\zeta_b\tau_y}\tau_x\left(\cos\eta_b+i\sin\eta_b\sigma_x\right)+{\rm h.c.}\Big]\no\\
&=&\tau_x+SM\rho_z\sigma_z+f_be^{i(\eta_b\rho_z\sigma_x-\zeta_b\tau_y)}\rho_x\tau_x,
\label{eq:Hintra}
\eea

\noi where the index $b$ in the respective coefficients denotes that these are evaluated for this value of distance, e.g., we set $f_b=f(b)$. The dimensionless variable $M=(\pi\nu_FJ)^{-1}$ controls the strength of the exchange coupling. Note that $S=\pm1$ keeps track of the global orientation freedom of the moments of the two identical dimers and we have $S_{1,3}=-S_{2,4}=S$. In the above we set $\rho_\pm=(\rho_x\pm i\rho_y)/2$ and used that $R_{12}=R_{34}=b$ and $\theta_{12}=\theta_{34}=\pi/2$.

We now derive the expression of the inter-dimer Hamiltonian. Hamiltonian elements in the dimer space are expressed using the Pauli matrices $\lambda_{x,y,z}$. Here, $\lambda_z=+1$ ($\lambda_z=-1$) corresponds to the left (right) dimer. For convenience, we also define $\lambda_\pm=(\lambda_x\pm i\lambda_y)/2$. Firstly, the two dimers mix with each other through ``ho\-ri\-zontal" couplings between ``1$\leftrightarrow$3" and ``2$\leftrightarrow$4", which lead to:
\bea
\hat{H}_{\rm horizontal}&=&\lambda_+f_ae^{-i\zeta_a\tau_y}\tau_x\big(\cos\eta_a+i\sin\eta_a\sigma_y\big)+{\rm h.c.}\no\\
&=&f_ae^{i(\eta_a\lambda_z\sigma_y-\zeta_a\tau_y)}\lambda_x\tau_x,
\label{eq:Hhorizontal}
\eea

\noi where we used that $R_{13}=R_{24}=a$ and $\theta_{13}=\theta_{24}=\pi$. Next, we evaluate the ``diagonal" coupling Hamiltonian which stems from mixing of ``1$\leftrightarrow$4" and ``2$\leftrightarrow$3", that is:
\bea
&&\hat{H}_{\rm diagonal}=\lambda_+\rho_+f_ce^{-i\zeta_c\tau_y}\tau_x\left(\cos\eta_c+i\sin\eta_c\sigma_ye^{-i\theta\sigma_z}\right)\no\\
&&+\lambda_+\rho_-f_ce^{-i\zeta_c\tau_y}\tau_x\left(\cos\eta_c+i\sin\eta_c\sigma_ye^{i\theta\sigma_z}\right)+{\rm h.c.}\,,
\label{eq:HDiagonal}
\eea

\noi that we obtained after taking into account that $R_{14}=R_{23}=\sqrt{a^2+b^2}\equiv c$, $\theta_{14}=\pi-\theta$ and $\theta_{23}=\pi+\theta$, where $\tan\theta=b/a$. By putting together the horizontal and dia\-gonal couplings we obtain the inter-dimer Hamiltonian:
\bea
\hat{H}_{\rm inter}=f_ae^{i(\eta_a\lambda_z\sigma_y-\zeta_a\tau_y)}\lambda_x\tau_x+
f_c\cos\eta_ce^{-i\zeta_c\tau_y}\lambda_x\rho_x\tau_x\no\\
-f_c\sin\eta_ce^{-i\zeta_c\tau_y}\big(\cos\theta\lambda_y\rho_x\tau_x\sigma_y+\sin\theta\lambda_x\rho_y\tau_x\sigma_x\big).\phd\label{eq:Hinter}
\eea

The total YSR Hamiltonian $\hat{H}_{\rm YSR}=\hat{H}_{\rm intra}+\hat{H}_{\rm inter}$ possesses a chiral symmetry effected by $\hat{\Pi}=\lambda_x\tau_y\sigma_x$, a generalized TRS symmetry effected by $\hat{\Theta}=\lambda_x\sigma_z\hat{{\cal K}}$, and a charge-conjugation symmetry induced by $\hat{\Xi}=\tau_y\sigma_y\hat{{\cal K}}$. The latter gua\-ran\-tees that zero energy eigenstates are of the PMM type. Due to their topologically trivial origin, these need to come in pairs. In the absence of additional symmetries, $\hat{H}_{\rm YSR}$ is in BDI symmetry class~\cite{Ryu}. However, $\hat{H}_{\rm YSR}$ also features the unitary mirror symmetry $y\mapsto-y$, effected by $\hat{\cal O}=\rho_x\sigma_y$. While this symmetry is not essential for the diode effect, it yet simplifies our ana\-lysis, as it allows us to block diagona\-li\-ze $\hat{H}_{\rm YSR}$. Note that this symmetry can be broken by assuming, for instance, different strengths $M\pm\delta M$ for the exchange couplings of the electrons to the two adatoms of each dimer. Such a mismatch can be added by the term $\hat{H}_{\rm mis}=S\delta M\sigma_z$.

Up to now, all
adatoms have been con\-si\-de\-red to feel the same superconducting phase. However, the Josephson effect of interest appears only for a nonzero phase bias $\phi$ between the two dimers. At this stage we incorporate the three different phases dictating the device in Fig.~\ref{fig:Figure1}(a) by twisting the YSR Hamiltonian in the following fashion:
\bea
&&\hat{H}_{\rm YSR}(\phi_{0,1,2})=e^{-i\phi_0\tau_z/2}\hat{H}_{\rm inter}\,e^{i\phi_0\tau_z/2}\\
&&+e^{-i\left(\phi_1\frac{1+\lambda_z}{2}+\phi_2\frac{1-\lambda_z}{2}\right)\tau_z/2}\hat{H}_{\rm intra}e^{i\left(\phi_1\frac{1+\lambda_z}{2}+\phi_2\frac{1-\lambda_z}{2}\right)\tau_z/2}.\no\label{eq:TwistedHamTheFirstTwist}
\eea

\noi To proceed, we perform a gauge transformation that removes the phase-twisting from the intra-dimer Hamiltonian and transfers it to the inter-dimer part. Besides this, we also replace the phases $\phi_{0,1,2}$ by the two independent phases $\phi$ and $\phi_{c0}=\phi_c-\phi_0$, where we introduced the ``center-of-mass" phase $\phi_c=(\phi_1+\phi_2)/2$. Given the above, the phase-twisted YSR Hamiltonian reads as:
\begin{align}
\hat{H}_{\rm YSR}(\phi,\phi_{c0})=\hat{H}_{\rm intra}+\hat{{\cal U}}(\phi,\phi_{c0})\hat{H}_{\rm inter}\,\hat{{\cal U}}^\dag(\phi,\phi_{c0})\,,\label{eq:TwistedHam}
\end{align}

\noi which is compactly expressed using the phase-twisting operator $\hat{{\cal U}}(\phi,\phi_{c0})={\rm Exp}[i\phi_{c0}\tau_z/2]{\rm Exp}[i\phi\lambda_z\tau_z/4]$. Notably, for $\phi_{c0}=\{0,\pi\}$, the Hamiltonian in Eq.~\eqref{eq:TwistedHam} preserves all the symmetries of its untwisted counterpart $\hat{H}_{\rm YSR}$. For all other values of $\phi_{c0}$, however, the phase-twisted Hamiltonian only respects $\hat{\cal O}$ and $\hat{\Xi}$ symmetries.

\textit{\bt{Restricting to physical states} -} To proceed, we note that not all of the eigenstates of $\hat{H}_{\rm YSR}(\phi)$ are physical. In fact, half of these do not lead to eigenenergies which are much smaller than unity, a condition which we postulated in order for the approximate form of the YSR couplings in Eq.~\eqref{eq:YSRcouplings} to be valid. Similar to Ref.~\onlinecite{Pientka}, we also project these states out from the very beginning of our analysis.

Since throughout this work the adatoms are assumed to be only weakly coupled, it is straightforward to identify the physical states with the eigenstates of the intra-dimer term $\tau_x+SM\rho_z\sigma_z\equiv\tau_x\big(1+SM\rho_z\tau_x\sigma_z\big)$, which yield eigenvalues $\pm(1-M)$. Specifically, for $\rho_z=\pm1$, we correspondingly keep the $\tau_x\sigma_z=\mp S$ eigenstates. Moreover, with an eye to the emergence of PMMs, we choose the projected basis states to simultaneously be eigenstates of $\hat{\Xi}$. The four states are written as column vectors in $\rho\otimes\tau\otimes\sigma$ space according to the following form:
\bea
\big|1;S\big>&=&\nicefrac{1}{2}\big(+i,\,+iS,\,-iS,\,+i,\,0,0,0,0\big)^\intercal\,,\\
\big|2;S\big>&=&
\nicefrac{1}{2}\big(-S,+1,+1,+S,0,0,0,0\big)^\intercal\,,\\
\big|3;S\big>&=&\nicefrac{1}{2}\big(0,\,0,\,0,\,0,\,+i,\,-iS,\,+iS,\,+i\big)^\intercal\,,\\
\big|4;S\big>&=&
\nicefrac{1}{2}\big(0,\,0,\,0,\,0,+S,+1,+1,-S\big)^\intercal\,,
\eea

\noi where $^\intercal$ denotes matrix transposition. By virtue of the choice of the above basis, we represent matrices in this space as Kronecker products of $\rho_{x,y,z}$ and $\kappa_{x,y,z}$ Pauli matrices along with their identity matrix companions.

Straightforward calculations lead to the projected phase-twisted Hamiltonian:
\bea
&&\hat{H}_{\rm YSR}'(\phi,\phi_{c0})=(1-M)\kappa_y-f_b\cos\zeta_b\sin\eta_b\rho_y\no\\
&&-Sf_b\sin\zeta_b\cos\eta_b\rho_y\kappa_x+f_a\cos\zeta_a\cos\eta_a\cos\phi_{c0}\lambda_x\kappa_y\no\\
&&+f_a\lambda_y\big[\sin\zeta_a\cos\eta_a\sin(\phi/2)-\sin\zeta_a\sin\eta_a\cos(\phi/2)\kappa_x\no\\
&&\qquad\,\,\quad-\cos\zeta_a\sin\eta_a\sin\phi_{c0}\kappa_z\big]\no\\
&&+f_c\sin\eta_c\big[S\cos\zeta_c\cos\theta\cos\phi_{c0}-\sin\zeta_c\sin\theta\sin(\phi/2)\big]\lambda_y\rho_y\kappa_y\no\\
&&-f_c\sin\eta_c\big[\cos\zeta_c\sin\theta\cos\phi_{c0}+S\sin\zeta_c\cos\theta\sin(\phi/2)\big]\lambda_x\rho_y\no\\
&&-Sf_c\cos\eta_c\lambda_x\rho_y\big[\sin\zeta_c\cos(\phi/2)\kappa_x+\cos\zeta_c\sin\phi_{c0}\kappa_z\big].
\label{eq:HamProjected}
\eea

\noi For $\phi_{c0}=\{0,\pi\}$, the above respects all the symmetries of the original Hamiltonian, but now with $\hat{\Pi}'=\lambda_x\rho_x\kappa_x$, $\hat{\Xi}'=\hat{\cal K}$, $\hat{\Theta}'=\hat{\Pi}'\hat{\cal K}$, and $\hat{\cal O}'=\rho_y$. The charge-conjugation symmetry ensures that the dispersions of the phase-twisted YSR states, which we here termed as YSR-ABS states, come in pairs, i.e., $\pm\varepsilon_s(\phi)$ where $s=1,2,3,4$. On top of that, the presence of mirror symmetry allows us to re\-pla\-ce $\rho_y$ by its two eigenvalues $\rho=\pm1$, and obtain two block Hamiltonians $\hat{H}_{{\rm YSR};\rho=\pm1}'$ which feature opposite eigenenergies. Hence, when mirror symmetry is present, we can associate $\varepsilon_{1,2,3,4}(\phi)$ with the four dispersions of the $\hat{H}_{{\rm YSR};\rho=+1}'$ block. When $\phi_{c0}\neq\{0,\pi\}$ only the $\Xi'$ and ${\cal O}'$ symmetries survive. Lastly, in the projected space, the moment mismatch enters via the term $\hat{H}_{\rm mis}'=-\delta M\rho_z\kappa_y$. This preserves only the charge-conjugation symmetry and enlists $\hat{H}_{\rm YSR}'(\phi)$ in class D.

The behavior of the YSR-ABS Josephson current is directly determined by the symmetry properties of the projected Hamiltonian and the associated dispersions $\varepsilon_s(\phi)$. The Josephson current of interest is the one flowing from the left to the right lead, which see a phase difference $\phi$. However, since here we are dealing with a three-terminal device, the desired current is also controlled by the value of $\phi_{c0}$. Depending on which two linear combinations of the phases $\phi_{0,1,2}$ are chosen to vary independently, $\phi_{c0}$ and $\phi$ may or may not be independent. Throughout this manuscript, we restrict to situations in which $\phi_{c0}$ can be kept fixed, while $\phi$ varies freely. Other phase configurations may also lead to an arbitrary periodicity for the Josephson current, as we briefly discuss in Ref.~\onlinecite{Sup}. Under the above assumptions, in the remainder we fix $\phi_{c0}=0$.

\textit{\bt{ Josephson current} -} To study the current it is preferrable to transfer to Fock space by introducing crea\-tion (annihilation) operators denoted $\hat{a}_s^\dag$ ($\hat{a}_s$) for the dia\-go\-na\-li\-zed YSR-ABS dispersions. These satisfy standard anticommutation relations $\big\{\hat{a}_s,\hat{a}_{s'}^\dag\big\}=\delta_{s,s'}\hat{\mathds{1}}$ and $\big\{\hat{a}_s,\hat{a}_{s'}\big\}=\hat{0}$. With the help of the above, we extend the YSR-BdG Hamiltonian to a many-body framework using the mapping $\hat{H}_{\rm YSR}'(\phi)\mapsto\hat{\cal H}_{\rm YSR}'(\phi)$, where we introduced:
\begin{align}
\hat{\cal H}_{\rm YSR}'(\phi)=\frac{1}{2}\sum_{s=1,2,3,4}\varepsilon_s(\phi)\big(2\hat{a}_s^\dag\hat{a}_s-1\big)\,.
\end{align}

We observe that the system's energy and the Josephson current depend on the fermion parities (FPs) of the ABSs. These are given by ${\cal P}_s(\phi)=2\big<\hat{a}_s^\dag\hat{a}_s\big>-1$ with $s=1,2,3,4$. Each FP takes the values $\pm1$ when the respective ABS state is occupied/empty. The precise value of each FP crucially depends on whether the system is in equilibrium or not~\cite{FuKaneJ,San-Jose2012,CarloJ}. In equilibrium, the system can exchange particles with its environment. As a result, the FP of the $s$-th ABS at zero temperature is given as ${\cal P}_s(\phi)=-{\rm sgn}\big[\varepsilon_s(\phi)\big]$. In contrast, when the occupancy of each ABS is fixed during the entire measurement process, one obtains a distinct nonequilibrium Josephson current for each FP configuration. In either case, the zero-temperature Josephson current is given by:
\begin{align}
I(\phi)=-\sum_s\frac{d\varepsilon_s(\phi)}{d\phi}\,{\cal P}_s(\phi),
\end{align}

\noi where we set the electric charge quantum $e$ equal to unity.

\begin{figure}[t!]
\begin{center}
\includegraphics[width=1\columnwidth]{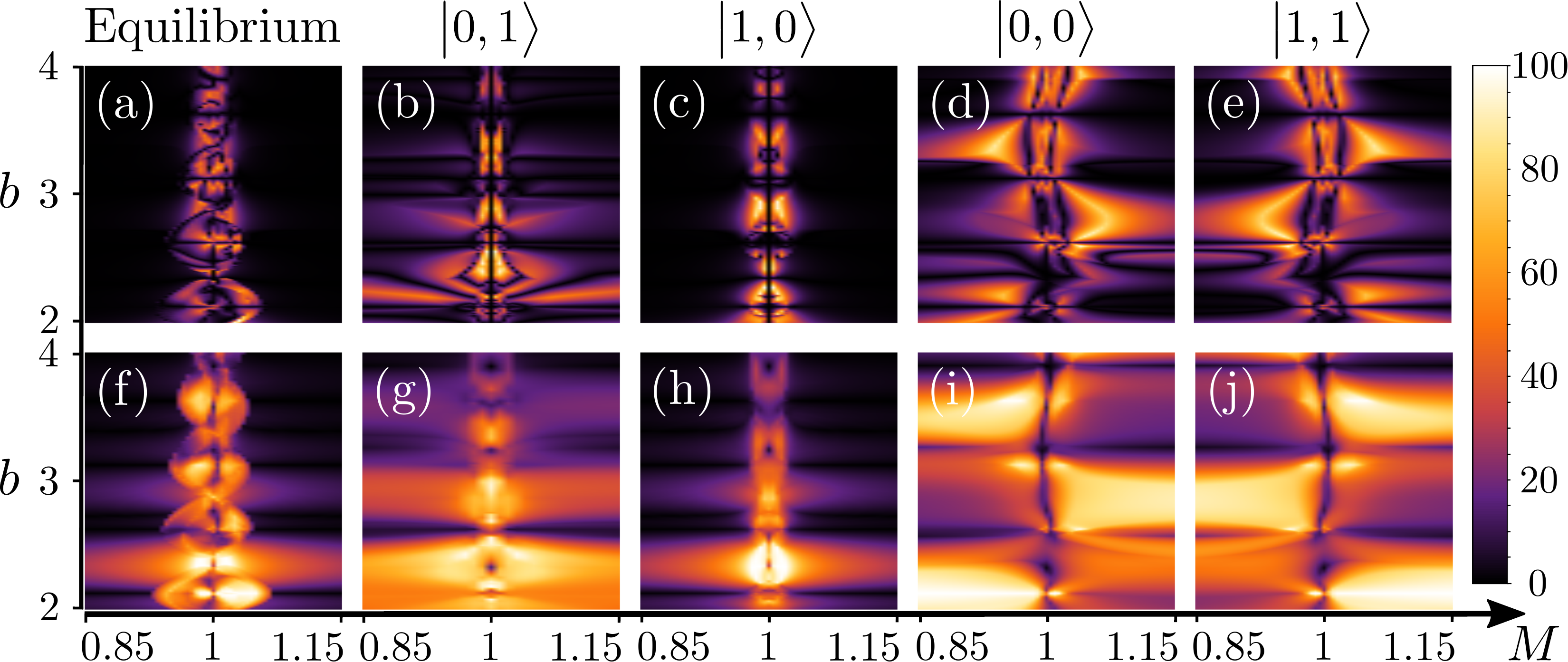}
\end{center}
\caption{Heat maps of the diode efficiency $Q[\%]$ (upper row) and the $4\pi$-pe\-rio\-di\-city indicator $P[\%]$ (lower row), as functions of the dimensionless moments strength $M$ and the intra-dimer adatom distance $b$ (in units of $\lambda_F$). We show results for both equilibrium and nonequilibrium cases. For the latter, we consider all four possibilities for the occupations $n_{2,3}=\{0,1\}$ of the two middle bands $\varepsilon_{2,3}(\phi)$. Here, the energy dispersions are obtained by diagonalizing $\hat{H}_{{\rm YSR};\rho=+1}'$. The nonequilibrium results are labelled according to a Fock space notation $\big|n_2,n_3\big>$. The resulting bands are substantially $4\pi$-periodic in an extended parameter regime, while the diode efficiency also reaches high values $\sim50\%$. In fact, a high degree of $4\pi$-periodicity is found in the even pa\-ri\-ty sector $\big\{\big|0,0\big>,\big|1,1\big>\big\}$. Moreover, the two distinct parity sectors exhibit a qualitatively different beha\-vior, thus giving rise to for a quantum Josephson diode effect. Lastly, we observe that the results for the equilibrium case are quite similar to the ones for $\big|1,0\big>$.}
\label{fig:Figure2}
\end{figure}

From Eq.~\eqref{eq:HamProjected} we observe that $\hat{H}_{\rm YSR}'(\phi,\phi_{c0}=0)$ and, in turn, the YSR-ABS dispersions and their Josephson current lack symmetry under the inversion of the phase bias $\phi\mapsto-\phi$. Such a property is required in order to obtain a nonreciprocal Josephson effect. Note additionally that $\hat{H}_{\rm YSR}'(\phi+2\pi,\phi_{c0}=0)\neq\hat{H}_{\rm YSR}'(\phi,\phi_{c0}=0)$ while $\hat{H}_{\rm YSR}'(\phi+4\pi,\phi_{c0}=0)=\hat{H}_{\rm YSR}'(\phi,\phi_{c0}=0)$. Hence, as it is typical for Hamiltonians describing coupled Majorana ope\-ra\-tors, the dependence of these couplings is $4\pi$- instead of $2\pi$-periodic in $\phi$~\cite{KitaevUnpaired}. We remark, however, that the $4\pi$-periodicity of the Hamiltonian terms does not always imply a similar periodicity for the Josephson current~\cite{FuKaneJ,San-Jose2012,CarloJ,KMC,MKC}. The structure of the Hamiltonian enables us to more transparently study such properties of the energy dispersions with the Pfaffian (Pf) of the real skew-symmetric matrix $\hat{B}(\phi)=i\hat{H}_{\rm YSR}'(\phi)$~\cite{KitaevUnpaired,Levchenko,KMC,Sakurai}, which satisfies the relation ${\rm Pf}[\hat{B}(\phi)]=\prod_{s=1,2,3,4}\varepsilon_s(\phi)$.

\textit{\bt{Numerical results} -} We now proceed with the exploration of the degree of nonreciprocity and $4\pi$-periodicity of the Josephson current. For this purpose, we introduce the two quantities (for $Q$ see also Refs.~\onlinecite{Daido2022prl,Ilic2022,He2022}):
\begin{align}
Q=\frac{|I_{\rm max}|-|I_{\rm min}|}{|I_{\rm max}|+|I_{\rm min}|}, P=\int_0^{2\pi}\frac{d\phi}{2\pi}\frac{\big|I(\phi)-I_{\rm shift}(\phi)\big|}{\big|I(\phi)\big|+\big|I_{\rm shift}(\phi)\big|+\Gamma},\no
\end{align}

\noi where $I_{{\rm max},{\rm min}}$ define the maximum and minimum va\-lues of the current $I(\phi)$, while $I_{\rm shift}(\phi)=I(\phi+2\pi)$. A small positive number $\Gamma=0^+$ is added to avoid sin\-gu\-la\-ri\-ties when $I(\phi)=I_{\rm shift}(\phi)=0$. We examine equilibrium and nonequilibrium scenarios. By assuming the hierar\-chy $\varepsilon_1(\phi)<\varepsilon_2(\phi)<\varepsilon_3(\phi)<\varepsilon_4(\phi)$, we explore four types of nonequilibrium currents. These are obtained by fi\-xing ${\cal P}_1(\phi)=+1$ and ${\cal P}_4(\phi)=-1$, while considering the four possible scenarios ${\cal P}_{2,3}(\phi)=\pm1$ for the FPs of $\varepsilon_{2,3}(\phi)$.

\begin{figure}[t!]
\begin{center}
\includegraphics[width=1\columnwidth]{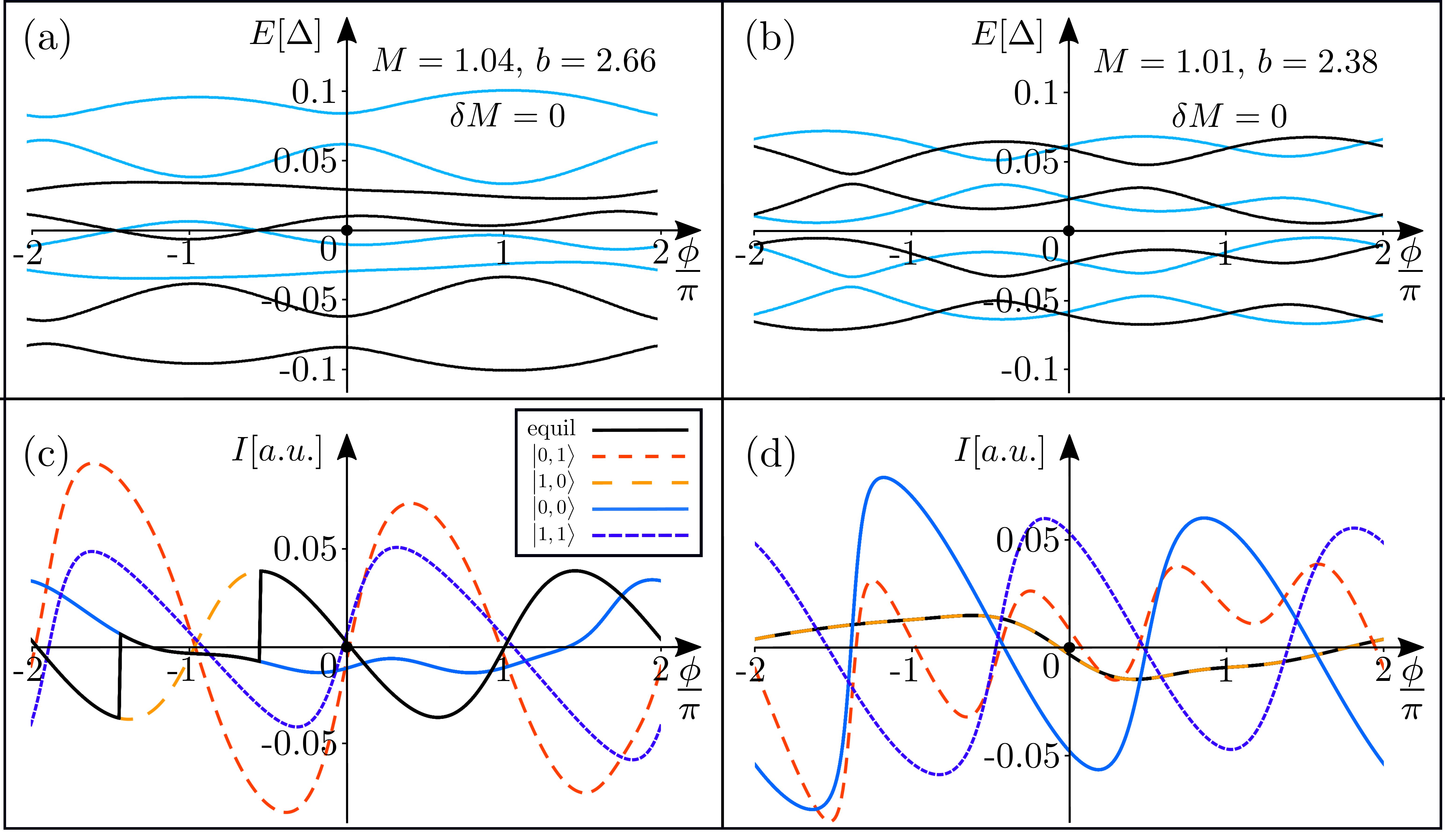}
\end{center}
\caption{Energy dispersions (a)-(b) and corresponding Josephson currents (c)-(d) for two different values for $b$ and $M$. In (a) the bands contain band crossings, while in (b) there is a full gap. In (c) and (d) we show results for the equilibrium and nonequilibrium currents. For the latter, we make use of the notation introduced in Fig.~\ref{fig:Figure2}. In (c) the maximum diode efficiency and strongest $4\pi$-periodicity are achieved for the even-parity sector state $\big|0,0\big>$, with $Q=40.5\%$ and $P=80.8\%$. The corresponding values obtained for the remaining states are much smaller. (d) shows results for the odd-parity state $\big|0,1\big>$, which supports both a substantially $4\pi$-periodic current and high diode efficiency with $Q=37.1\%$ and $P=67.4\%$.}
\label{fig:Figure3}
\end{figure}

Outcomes from our numerics for $S=1$ are shown in Figs.~\ref{fig:Figure2} and~\ref{fig:Figure3}. From the former, we confirm the nonreciprocal and $4\pi$-periodic nature of the current. Note, however, that a strong $4\pi$-periodicity does not necessa\-ri\-ly imply a high diode efficiency, since the $4\pi$-periodicity is fully compatible with the inversion operation $\phi\mapsto-\phi$. A strong diode effect is obtained in a window centered around $M=1$, which is precisely the value at which PMMs emerge in the absence of the phase bias. As a matter of fact, from Figs.~\ref{fig:Figure1}(b) and~\ref{fig:Figure3} we find that the near-zero YSR-ABS branches, i.e., the ones descending from PMMs, are highly asymmetric and are the main contri\-bu\-tors to the diode effect. In the PMM regime $M\sim1$, the term $(1-M)\kappa_y$ becomes small and comparable to the inter-adatom couplings, thus giving rise to the complex interplay shown in Fig.~\ref{fig:Figure2}. Interestingly, the results for the nonequilibrium states are mirrored about $M=1$. This is due to the phase-independence of the nonequilibrium FPs and a symmetry relation of $\hat{H}_{{\rm YSR};\rho=+1}'(\phi,\phi_{c0}=0)$ under $(1-M,\phi)\mapsto (M-1,2\pi-\phi)$. On the other hand, a common feature for both equilibrium and nonequilibrium states is the diminishing of the diode efficiency upon increasing $b$, as well as the arising semi-periodic dependence of $Q$ and $P$ on this parameter. This feature results from the phase factor in Eq.~\eqref{eq:Hintra} which involves $\zeta_b$ and controls the alternating conversion of electron-to-pairing hopping and vice versa. This semi-periodicity arises when $b$ is approximately equal to an integer multiple of $\lambda_F/2$.

In Fig.~\ref{fig:Figure3}, we show the current for two cases with an equally high diode effect$\sim40\%$, while supporting gapless and gapped band structures. From this, we infer that band (crossings) touchings in (non)equilibrium cases tend to enhance both $Q$ and $P$. It is important to point out that for the here-assumed hierarchy of the dispersions, we expect the equilibrium and nonequlibrium $\big|1,0\big>$ cases to give identical results when $\varepsilon_2(\phi)$ is occupied and $\varepsilon_3(\phi)$ is empty. This scenario takes place in Fig.~\ref{fig:Figure3}(d). In Fig.~\ref{fig:Figure3}(c), however, the equilibrium and $\big|1,0\big>$ states yield different outcomes, which is attributed to the presence of the band crossing. This is because the equilibrium FP is set by which state is positive/negative, while for the $\big|1,0\big>$ configuration the FP is fixed for all values of $\phi$.

\textit{\bt{Discussion} -} We reveal that the Josephson effect in the vicinity of PMMs is equally rich to that of MZMs. Se\-ve\-ral recent works~\cite{LeggMZM,Cayao2023,Liu2023} have discussed the impact of MZMs on the diode effect. Reference~\onlinecite{LeggMZM} exa\-mi\-ned the robustness of the effect against FP nonconservation, while the other two works restricted to equilibrium response. Here, we do not only prove that dispersions associated to PMMs can lead to a substantial diode effect with efficiencies rea\-ching $50\%$, but we also show that the Josephson current can be $4\pi$-periodic even in equilibrium. In fact, any value for the periodicity of the Josephson current is in principle attainable in the setup shown in Fig.~\ref{fig:Figure1}(a)~\cite{Sup}. Moreover, we find a quantum Josephson effect which depends on the FPs of the ABS dispersions.  Notably, we expect the high efficiencies obtained here to be robust against weak disorder. For instance, we find that introducing a small moment mismatch
$\delta M=0.015$ to the case of Fig.~\ref{fig:Figure3}(a), results to a non-substantially modified band structure, that we depict in Fig.~\ref{fig:Figure1}(b).

In actual experiments, the Josephson current stemming from the YSR and PMM states pinned by the adatoms needs to be isolated from the total current by subtracting the contribution of the usual ABSs, which also becomes relevant in the mesoscopic device proposed here. The usual ABS contribution can be inferred by investigating an identical junction in the absence of the adatoms. Noteworthy, the Josephson current  stemming from the adatoms sa\-ti\-sfies the relation $I(-S,\phi)=I(S,-\phi)$, which is a property that can be exploited for its detection,  since the standard ABS contribution to the current does not follow this rule. An alternative pathway is to consider interferometric setups in which one arm contains a pair of dimers and the other does not, or, it contains a pair of dimers with inverted spin moment. Introducing a relative phase bias between the two arms enables us to tailor the diode effect, as well as to detect or control the underlying YSR contribution.

Concluding, it is important to emphasize that a key advantage of our diode proposal is that it is free from the requirement of topological superconductivity, which makes it pro\-mi\-sing to be realized in currently accessible platforms. First, our ideas can find applicability in already existing Majorana platforms consisting of magnetic chains embedded on Rashba SCs~\cite{Yazdani1,Yazdani2,Kim2018,Scheider2020,Schneider2021,Schneider2023}. Notably, Ref.~\onlinecite{Schneider2023} already discusses AFM chains and, thus, it appears prominent for realizing the configuration in Fig.~\ref{fig:Figure1}(a), and potentially addressing the adatoms by employing superconducting STM tips. The other possible class of systems concerns coupled quantum dots with induced superconductivity~\cite{Grove2018,Saldana2018,Saldana2020,Dvir2023,Zatelli,tenHaaf}, which are purposed for engineering a Kitaev chain~\cite{Sau2012,Fulga,CXLiu2023}. Note that Refs.~\onlinecite{Dvir2023,Zatelli} have already produced a PMM dimer. Such platforms appear ideal to realize our proposal since the standard Josephson current contribution is absent. We hope that our work will motivate theorists and experimenta\-lists alike to carry out dedicated studies in these systems.

\section*{Acknowledgements}

We are thankful to M. Cuoco, K. Flensberg, and M.~T. Mercaldo for motivating and helpful discussions. M.~R. acknow\-ledges support from the Novo Nordisk Foundation grant
NNF20OC0060019.

\appendix

\begin{widetext}

\section*{\protect\huge Supplemental Material}

\section*{ Nonreciprocal Equilibrium Josephson Effect of Arbitrary Periodicity\\  from Poor Man's Majorana Zero Modes}

\subsection*{Panagiotis Kotetes$^{1,2}$, Merc\`e Roig$^3$, and Brian M. Andersen$^3$}

\begin{center}
$^1$ Beijing Academy of Quantum Information Sciences, Beijing 100193, China\\
$^2$ CAS Key Laboratory of Theoretical Physics, Institute of Theoretical Physics,\\ Chinese Academy of Sciences, Beijing 100190, China\\
$^3$ Niels Bohr Institute, University of Copenhagen, DK-2100 Copenhagen, Denmark
\end{center}

\section*{I. Additional Results upon Varying\\ the Pairing Gap and the Strength of the Spin-Orbit Interaction}

In this section we provide additional results extending those presented in Fig.~3 of the main text, by considering smaller values for the ratios $\Delta/E_F$ and $2m_*\upsilon/p_F$. Specifically, we now set $\Delta/E_F=0.001$ and examine two cases for the strength of the Rashba spin-orbit interaction (SOI), that is, we consider $2m_*\upsilon/p_F=0.1$ as in the main text and $2m_*\upsilon/p_F=0.01$. These are presented in the figures below.

\begin{figure*}[h!]
\begin{center}
\includegraphics[width=1\textwidth]{Figure_Sup_1}
\end{center}
\caption{Additional calculations in the spirit of the ones presented in Fig.~3 of the main text obtained for $\delta M=0$, $\Delta/E_F=0.001$, and two different values for the parameter $2m_*\upsilon/p_F$ quantifying the SOI strength. The diode efficiency is somehow reduced upon solely reducing the ratio $\Delta/E_F$. Reducing instead the SOI strength leads to a drastic suppression of the diode effect.}
\label{fig:Figure_Sup_1}
\end{figure*}

These additional numerical findings show that the diode effect persists as we decrease the ratio $\Delta/E_F$ while keeping the values for the parameter $M$ and the SOI strength unchanged. The main notable differences are that now the YSR states appear at higher energies (in units of $\Delta$) and the associated Josephson diode effect becomes reduced. Indeed, for $M=1.04$ and $b=2.66$ we find now that the maximum efficiency is obtained in the equilibrium case with $Q=15.1\%$ and $P=40.2\%$, while for  $M=1.01$ and $b=2.38$ the maximum diode efficiency is obtained for the nonequilibrium configuration  $|0,1\rangle$ with $Q=8.8\%$ and $P\simeq79\%$. In contrast, decreasing the SOI strength leads to the drastic diminishing of the diode effect since it renders the energy dispersions featureless. Indeed, when evaluating  for $\Delta/E_F=0.001$ and $2m_*\upsilon/p_F=0.01$, we find that for $M=1.04$ and $b=2.66$ the maximum efficiency is obtained in the $|0,0\rangle$  case with $Q=4.9\%$ and $P=84\%$, while for $M=1.01$ and $b=2.38$ the maximum diode efficiency is obtained for the nonequilibrium configuration  with $|0,1\rangle$ with $Q=2.35\%$ and $P\simeq10\%$. We observe that when both  $\Delta/E_F$ and $2m_*\upsilon/p_F$ are reduced, the fermion parity of the state for which the maximum efficiency occurs remains unaltered. In any case, the above prove that the diode effect persists for a variety of parameter values with, however, a strong effect obtained for the values chosen in the manuscript.

\section*{II. Additional Results for Alternative Periodicities}

In this section we carry out additional calculations of the Josephson diode effect for other cases in which the periodicity differs from $4\pi$. In particular, we append below results for the cases $\phi_{c0}=\phi/2$ and $\phi_{c0}=\phi/3$ where the current becomes $2\pi$- and $12\pi$-periodic, respectively. For the former case we also show results for both $S=\pm1$ values in order to numerically verify that $I(-S;\phi)=I(S;-\phi)$, while in the latter case the $4\pi$-indicator is replaced by a $12\pi$-indicator. For this purpose, we replace $I_{\rm shift}(\phi)=I(\phi+2\pi)$ by $I_{\rm shift}(\phi)=I(\phi+6\pi)$ in the definition of $P$.

\begin{figure*}[h!]
\begin{center}
\includegraphics[width=\textwidth]{Figure_Sup_2}
\end{center}
\caption{Additional results to the ones presented in Fig.~3 of the main text, for the parameter values $\delta M=0$, $\Delta/E_F=0.1$, and $2m_*\upsilon/p_F=0.1$, along with the phase configuration $\phi_{c0}=\phi/2$. The dispersions and currents are $2\pi$-periodic and become inverted according to $\phi\mapsto-\phi$, upon taking $S\mapsto-S$.}
\label{fig:Figure_Sup_2}
\end{figure*}

\begin{figure*}[h!]
\begin{center}
\includegraphics[width=0.64\textwidth]{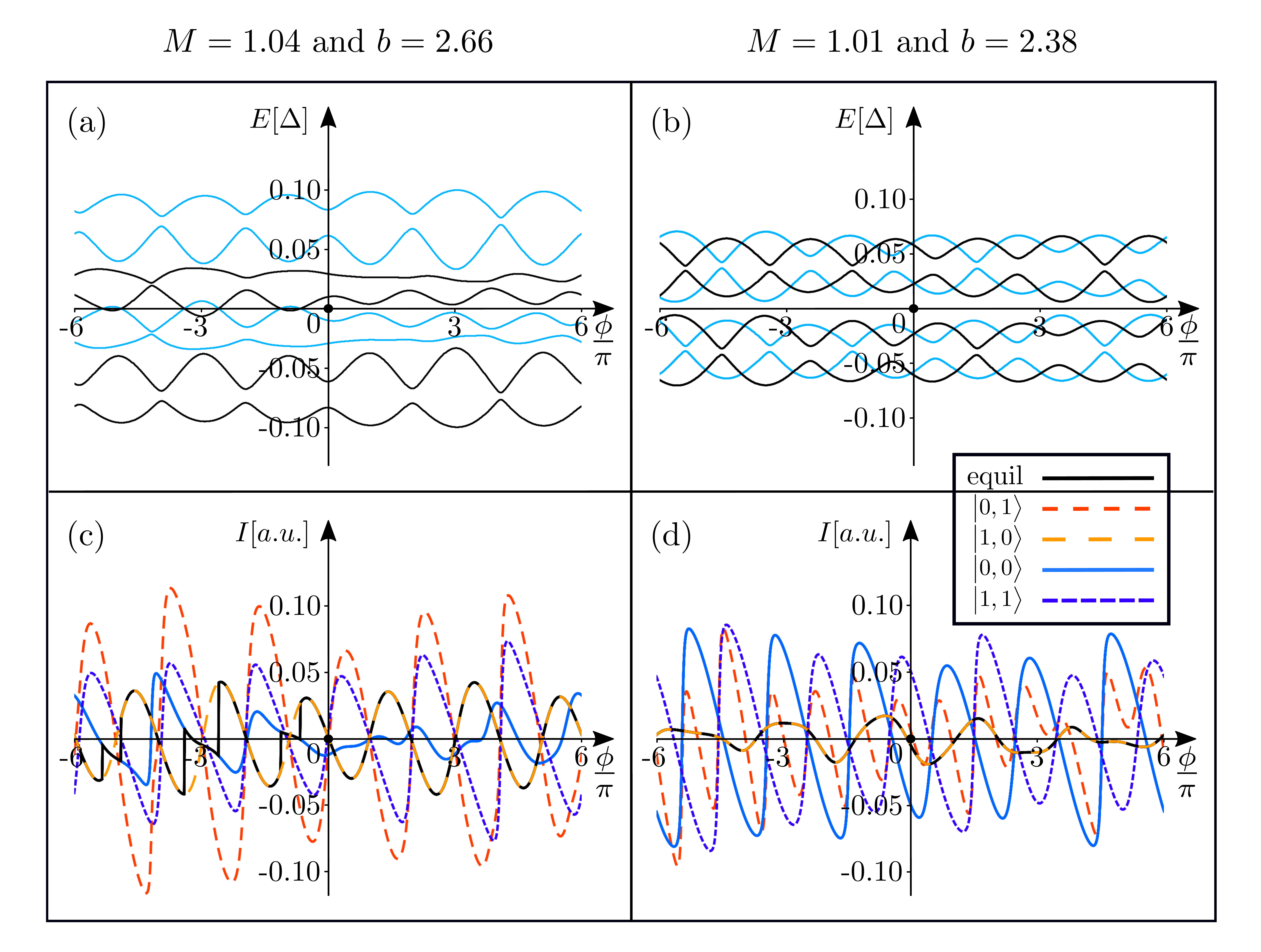}
\end{center}
\caption{Additional results to the ones presented in Fig.~3 of the main text, for the parameter values $\delta M=0$, $\Delta/E_F=0.1$, and $2m_*\upsilon/p_F=0.1$, along with the phase configuration $\phi_{c0}=\phi/3$. Here, the dispersions and currents are $12\pi$-periodic.}
\label{fig:Figure_Sup_3}
\end{figure*}

From the above results, we indeed confirm that $I(-S;\phi)=I(S;-\phi)$ and that the dispersions for $\phi_{c0}=\phi/2$ are $2\pi$-periodic. For $M=1.04$ and $b=2.66$ we find that the maximum efficiency is obtained in the $|0,0\rangle$ case with $Q=9.1\%$, while for $M=1.01$ and $b=2.38$ the maximum diode efficiency is obtained for the nonequilibrium configuration $|1,1\rangle$  with $Q=6.5\%$. Hence, we find changes in the maximum-efficiency state and a reduced diode efficiency compared to the $4\pi$-periodic case considered in the main text.

As shown instead in the results of Fig.~\ref{fig:Figure_Sup_3} of this supplementary section, in the case where we assume the phase dependence $\phi_{c0}=\phi/3$, we obtain that when we consider $M=1.04$ and $b=2.66$ the maximum efficiency is given by the nonequilibrium case $|0,0\rangle$ with $Q=17.6\%$ and with the $12\pi$-periodicity indicator $P=81.7\%$, while for $M=1.01$ and $b=2.38$ the maximum diode efficiency is obtained for the nonequilibrium configuration  with $|0,1\rangle$ and $Q=7.3\%$ with the $12\pi$-periodicity indicator $P\simeq54\%$. From the above, we find that the diode effect deteriorates when departing from the $4\pi$-periodicity, while the case $\phi_{c0}=\phi/3$ is more similar to the $4\pi$-periodic one discussed in the main text than the one with $\phi_{c0}=\phi/2$.

\end{widetext}

\end{document}